\newbox\leftpage
\newdimen\fullhsize
\newdimen\hstitle
\newdimen\hsbody
\tolerance=1000\hfuzz=2pt
\overfullrule=0pt
\def\bigans{b }
\ifx\answ\bigans\message{(this will come out unreduced.}


\magnification=1200\baselineskip=20pt
\font\titlefnt=amr10 scaled\magstep3\global\let\absfnt=\tenrm
\font\titlemfnt=ammi10 scaled\magstep3\global\let\absmfnt=\teni
\font\titlesfnt=amsy10 scaled\magstep3\global\let\abssfnt=\tensy
\hsbody=\hsize \hstitle=\hsize 
 
\else\def\apans{h }
\message{(this will be reduced.}
\let\lr=l
\magnification=1000\baselineskip=12pt\voffset=-.31truein\hoffset=-.59truein
\hstitle=7truein\hsbody=3.4truein\vsize=9.5truein\fullhsize=7truein
\ifx\apansw\apans\special{ps: landscape}\hoffset=-.54truein
  \else\voffset=-.25truein\hoffset=-.45truein\fi
 
\output={\ifnum\count0=1 
  \shipout\vbox{\hbox to \fullhsize{\hfill\pagebody\hfill}}\advancepageno
  \else
  \almostshipout{\leftline{\vbox{\pagebody\makefootline}}}\advancepageno
  \fi}
\def\almostshipout#1{\if l\lr \count1=1
      \global\setbox\leftpage=#1 \global\let\lr=r
   \else \count1=2
      \shipout\vbox{\hbox to\fullhsize{\box\leftpage\hfil#1}}
      \global\let\lr=l\fi}
\fi

%

\def\draftmode{\message{ DRAFTMODE }\def\draftdate{{\rm preliminary draft:
\number\month/\number\day/\number\yearltd\ \ \hourmin}}%
\headline={\hfil\draftdate}\writelabels\baselineskip=20pt plus 2pt minus 2pt
 {\count255=\time\divide\count255 by 60 \xdef\hourmin{\number\count255}
  \multiply\count255 by-60\advance\count255 by\time
  \xdef\hourmin{\hourmin:\ifnum\count255<10 0\fi\the\count255}}}

\def\title#1#2{\nopagenumbers\absfnt\hsize=\hstitle\rightline{}%
\centerline{\titlefnt\textfont0=\titlefnt%
\textfont1=\titlemfnt\textfont2=\titlesfnt #1}%
\centerline{\titlefnt\textfont0=\titlefnt%
\textfont1=\titlemfnt\textfont2=\titlesfnt #2
}%
\textfont0=\absfnt\textfont1=\absmfnt\textfont2=\abssfnt\vskip .5in}
 
\def\date#1{\vfill\leftline{#1}%
\tenrm\textfont0=\tenrm\textfont1=\teni\textfont2=\tensy%
\supereject\global\hsize=\hsbody%
\footline={\hss\tenrm\folio\hss}}
%

\def\nolabels{\def\eqnlabel##1{}\def\eqlabel##1{}\def\reflabel##1{}}
\def\writelabels{\def\eqnlabel##1{\hfill\rlap{\hskip.09in\string##1}}%
\def\eqlabel##1{\rlap{\hskip.09in\string##1}}%
\def\reflabel##1{\noexpand\llap{\string\string\string##1\hskip.31in}}}
\nolabels
%
\global\newcount\secno \global\secno=0
\global\newcount\meqno \global\meqno=1
 
\def\newsec#1{\global\advance\secno by1\xdef\secsym{\the\secno.}
\global\meqno=1
\bigbreak\bigskip
\noindent{\bf\the\secno. #1}\par\nobreak\medskip\nobreak}
\xdef\secsym{}
 
\def\appendix#1#2{\global\meqno=1\xdef\secsym{#1.}\bigbreak\bigskip
\noindent{\bf Appendix #1. #2}\par\nobreak\medskip\nobreak}
 
 
\def\eqnn#1{\xdef #1{(\secsym\the\meqno)}%
\global\advance\meqno by1\eqnlabel#1}
\def\eqna#1{\xdef #1##1{(\secsym\the\meqno##1)}%
\global\advance\meqno by1\eqnlabel{#1$\{\}$}}
\def\eqn#1#2{\xdef #1{(\secsym\the\meqno)}\global\advance\meqno by1%
$$#2\eqno#1\eqlabel#1$$}
 
\global\newcount\ftno \global\ftno=1
\def\refsymbol{\ifcase\ftno
\or\dagger\or\ddagger\or\P\or\S\or\#\or @\or\ast\or\$\or\flat\or\natural
\or\sharp\or\forall
\or\oplus\or\ominus\or\otimes\or\oslash\or\amalg\or\diamond\or\triangle
\or a\or b \or c\or d\or e\or f\or g\or h\or i\or i\or j\or k\or l
\or m\or n\or p\or q\or s\or t\or u\or v\or w\or x \or y\or z\fi}
\def\foot#1{{\baselineskip=14pt\footnote{$^{\refsymbol}$}{#1}}\ %
\global\advance\ftno by1}

 
\global\newcount\refno \global\refno=1
\newwrite\rfile
\def\ref#1#2{${[\the\refno]}$\nref#1{#2}}
\def\nref#1#2{\xdef#1{${[\the\refno]}$}%
\ifnum\refno=1\immediate\openout\rfile=refs.tmp\fi%
\immediate\write\rfile{\noexpand\item{\the\refno.\ }\reflabel{#1}#2.}%
\global\advance\refno by1}
\def\addref#1{\immediate\write\rfile{\noexpand\item{}#1}}
 
\def\semi{;\hfil\noexpand\break}




\hyphenation{anom-aly anom-alies coun-ter-term coun-ter-terms}
 


 

\def\pmb#1{\setbox0=\hbox{#1}%
 \kern-.025em\copy0\kern-\wd0
 \kern .05em\copy0\kern-\wd0
 \kern-.025em\raise.0433em\box0 }



\def\cp #1 #2 #3 {{\sl Chem.\ Phys.} {\bf #1}, #2 (#3)}
\def\jetp #1 #2 #3 {{\sl Sov.\ Phys.\ JETP} {\bf #1}, #2 (#3)}
\def\jfm #1 #2 #3 {{\sl J. Fluid\ Mech.} {\bf #1}, #2 (#3)}
\def\jpa #1 #2 #3 {{\sl J. Phys.\ A} {\bf #1}, #2 (#3)}
\def\jcp #1 #2 #3 {{\sl J.\ Chem.\ Phys.} {\bf #1}, #2 (#3)}
\def\jpc #1 #2 #3 {{\sl J.\ Phys.\ Chem.} {\bf #1}, #2 (#3)}
\def\jsp #1 #2 #3 {{\sl J.\ Stat.\ Phys.} {\bf #1}, #2 (#3)}
\def\jdep #1 #2 #3 {{\sl J.\ de Physique I} {\bf #1}, #2 (#3)}
\def\macromol #1 #2 #3 {{\sl Macromolecules} {\bf #1}, #2 (#3)}
\def\pra #1 #2 #3 {{\sl Phys.\ Rev.\ A} {\bf #1}, #2 (#3)}
\def\prb #1 #2 #3 {{\sl Phys.\ Rev.\ B} {\bf #1}, #2 (#3)}
\def\pre #1 #2 #3 {{\sl Phys.\ Rev.\ E} {\bf #1}, #2 (#3)}
\def\prl #1 #2 #3 {{\sl Phys.\ Rev.\ Lett.} {\bf #1}, #2 (#3)}
\def\prsl #1 #2 #3 {{\sl Proc.\ Roy.\ Soc.\ London Ser. A} {\bf #1}, #2 (#3)}
\def\rmp #1 #2 #3 {{\sl Rev.\ Mod.\ Phys.} {\bf #1}, #2 (#3)}
\def\zpc #1 #2 #3 {{\sl Z. Phys.\ Chem.} {\bf #1}, #2 (#3)}
\def\zw #1 #2 #3 {{\sl Z. Wahrsch.\ verw.\ Gebiete} {\bf #1}, #2 (#3)}

\def\sig{\sigma}

\def\halfspace{\hskip0.4cm}
\font\titlefont=cmbx10 scaled\magstep2


\centerline{\titlefont{Nonequilibrium brittle fracture propagation:}}
\centerline{\titlefont{Steady state, oscillations and intermittency}}
\bigskip
\bigskip
\centerline{{\bf Raphael Blumenfeld}}

\smallskip
\centerline{Theoretical Division and Center for Nonlinear Studies}
\centerline{Los Alamos National Laboratory, Los Alamos, NM 87545, USA}

\bigskip
\item{}{\bf Abstract}
\smallskip
A minimal model is constructed for two-dimensional fracture propagation. 
The heterogeneous process zone is presumed to suppress stress
relaxation rate, leading to non-quasistatic behavior. Using the
Yoffe solution, I construct and solve a dynamical equation for the tip
stress. I discuss a generic tip velocity response to local stress and find 
that noise-free propagation is either at steady state or oscillatory,
depending only on one material parameter.
Noise gives rise to intermittency and quasi-periodicity. The theory 
explains the velocity oscillations and the complicated behavior seen 
in polymeric and amorphous brittle materials. I suggest experimental
verifications and new connections between velocity measurements and 
material properties.


\date{LA-UR 94-3527}
\eject

The dynamics of cracks propagating in amorphous brittle media focused 
extensive study since the forties, mostly through quasi-static approaches 
and energetic arguments. In spite of recently renewed interest there are 
several fundamental issues that seem difficult to resolve in any simple 
way. For example: The limiting crack velocity, predicted to be the 
Rayleigh wave speed (RWS) in the bulk \ref\mott{N.F.Mott, Engineering 
{\bf 165}, 16 (1948); A.N.Stroh, Adv. Phys. {\bf 6}, 418 (1957)}, is 
observed to be only about half of that; The mechanism
for crack initiation and arrest is poorly understood; And the occurence 
of velocity oscillations \ref\osc{J.Fineberg, S.P.Gross, M.Marder and 
H.L.Swinney,
Phys. rev. Lett. {\bf 67}, 457 (1991); Phys. Rev. {\bf B45}, 5146 (1992);
J.Dear and H.MacGillivray, in Dynamic Failure of Materials, eds. 
H.P.Rosmanith and A.J.Rosakis (Elsevier, London 1991); W.G.Knauss and 
P.D.Washabaugh, Int. J. Fract., 189 (1993)} is still a puzzle. At the 
heart of the problem is the fact that the system's behavior depends on 
the lengthscale. While it is evident that the atomistic behavior differs 
from the continuous, it is this author's opinion that even on the 
continuum scale the physics near the tip is distinct from that far away 
and therefore should be treated differently. This may explain an
apparent discrepancy: On the one hand, since the bulk shear wave speed 
(SWS) is higher than the
crack velocity, it is clear that far from the crack quasi-static arguments
should work well because the field relaxes to its static form,
$\sig_{\alpha\beta}=K f_{\alpha\beta}(\theta)/\sqrt{2\pi r}$, sufficiently
fast. Here $\sig_{\alpha\beta}$ is the stress tensor, $r$ is the distance 
from the crack tip, $f_{\alpha\beta}$ depends only on the azimuthal angle 
$\theta$ and $K$ is the (time-dependent) stress intensity factor. On the 
other hand, the inability of quasi-static theories to account for the 
above phenomena suggests that much of the behavior is determined by the 
{\it local dynamics at the tip}
and hence that the propagation is a far-from-equilibrium process, 
indescribable by approaches that appeal to energy balancing. In the two 
scale picture the nonequilibrium dynamics act in effect to dress the tip 
singularity as seen from afar. The matching of the near and far fields at 
the crossover scale then yields the far-away behavior of $K$. A reasonable 
guess would be that the crossover scale is of the order of the size of the 
processing zone (PZ) in front of the propagating crack. While the far 
quasi-static field is well understood within linear elasticity, there is 
little understanding of the short range physics although a few 
phenomenological dynamic equations have been advanced 
\ref\lang{J.S.Langer, Phys. Rev. Lett. {\bf 70}, 3592 (1993); 
D.A.Kurtze and D.C.Hong, {\it ibid}. {\bf 71}, 847 (1993)} 
\ref\web{T.W.Webb and E.C.Aifantis, Int. J. of Solids Structures {\bf 32},
2725 (1995) and references therein}\ to explain the limiting tip velocity.

It has been conjectured \ref\rki{K.Ravi-Chandar and W.G.Knauss, Int. J. 
Fract. {\bf 25}, 247 (1984) and references therein; {\it ibid}. {\bf 26}, 65
(1984)}\ref\wk{P.D.Washabaugh and W.G.Knauss, Int. J. Fract. {\bf 65}, 97
(1994)} that the reason for the complicated short-range behavior is the
heterogeneous and fluid structure of the processing zone PZ. This conjecture
may be supported by observations of extremely slow relaxation rates of the
stress at the tip after arrest \ref\mf{L.B.Freund, in Proc. of the Workshop 
on Dynamic Fracture, eds. W.G.Knauss, K. Ravi-Chandar and A.J.Rosakis, 
CALTECH, (1983); C.C.Ma and L.B. Freund, Brown University Report (September 
1984)}, rates that are an order of magnitude below expectation had the 
relaxation taken place at the bulk speed of sound. This implies that the 
waves that re-establish the stress field in the PZ travel at a speed, $c$, 
that is {\it much lower} than the bulk SWS, probably due to scattering from 
microvoids.

The model proposed here concerns the short-range dynamics and takes on board
several ingredients: the low value of $c$ in the PZ, the occurrence of
different stresses for crack initiation and arrest \rki\ref\hyst{W.B.Bradley
and A.S.Kobayashi, Eng. Fract. Mech. {\bf 3}, 317 (1971); R.C.Hoagland et al.,
Metallurgical Trans. {\bf 3}, 123 (1972); L.B.Freund, in Mechanics of 
Fracture, ed. F.Erdogan, AMD Vol. 19, ASME, NY (1976); M.F.Kanninen, J. 
Mech. Phys. Sol., {\bf 25}, 69 (1977); P.B.Crosely and E.J.Ripling, 
ASTM-STP {\bf 627}, 372 (1977)}, and, based on existing observations, an 
assumed velocity response to the tip stress. These suffice to
construct and solve a dynamic equation. The explicit form of the
velocity-stress relation is not required for most of the results obtained 
here, only its qualitative behavior. The model leads to either a 
steady-state propagation at a limiting velocity or an oscillatory behavior, 
with the selection between the two modes depending on the location of the 
suppressed speed $c$ on the velocity response function. Introduction of 
noise due to microvoid distribution is shown to give rise to an 
intermittent propagation that can turn into a quasiperiodic behavior.

Consider a line crack (not necessarily straight) in a two-dimensional 
material. The PZ in front of the tip is modelled as an effective continuous 
medium with a reduced SWS, $c$. The dependence of the crack dynamics on 
material properties enters through a velocity response function, 
$v(\sig)$ \ref\fti{For simplicity I ignore possible dependence of $v$ on 
other parameters}, where $\sig$ is the local stress at the tip in the 
forward direction. As the crack propagates, the field near the tip adjusts 
to the changing
boundary at a rate that corresponds to $c$. Observations that steady state
propagation is at about half the bulk RWS, combined with the fact that $c$ is
much lower than the bulk (homogeneous) RWS, implies that the tip velocity can
{\it momentarily exceed} the local value of $c$. This is a basic assumption in
what follows. I comment that this does not violate the energy-balance which
holds for scales away from the PZ because near the tip the dynamic response,
$v(\sig)$, is swifter than the global energy equilibration process.
The measured behavior of $v(\sig)$ is hysteretic with two material-dependent
thresholds: $\sig_h$, above which propagation initiates, and $\sig_l<\sig_h$,
to which the stress has to drop for the crack to arrest \rki\hyst. For
$\sig>\sig_h$ the velocity is also known to increase very slowly with
stress\ref\df{E.g., J.W.Dally, W.L. Fourney and G.R.Irwin, Int. J. of Fract.,
159 (1985), and references therein} \ref\commi{Although most existing velocity
measurements are in terms of $K$, one can deduce the dependence on $\sigma$
through the relations in the text}. Fig. 1 shows a qualitative form of
$v(\sigma)$ that is consistent with experimental observations. This local 
non-monotonic response differs from that in \web which depends on energy 
eqilibration far from the tip. Its locality allows to find the dynamics 
without further assumptions. A local two-branch velocity can be derived 
from atomistic models \ref\thom{R.Thomson, in {\it Solid State Physics} 
ed. H. Ehrenreich {\it et al}. (Academic, NY, 1986) Vol. 39, p. 1; 
M.Marder, Phys. Rev. Lett. {\bf 74}, 4547 (1995); B.L.Holian, R.Blumenfeld 
and P.Gumbsch, in preparation}.

To derive the equation of motion of the tip, let us start from the Yoffe
solution for the forward field of a propagating crack of length $a$
\ref\yoffe{E.H.Yoffe, Phil. Mag. {\bf 42}, 739 (1951)},
\eqn\Iii{\sig = \sig_\infty\left[(\zeta+a)/\sqrt{\zeta(\zeta+2a)}\right]\ ,}
where $\zeta$ is the distance from the tip and $\sig_\infty$ is the tensile
stress applied perpendicular to the propagation axis far away from the crack.
In what follows the stress is measured in units of $\sig_\infty$ and
$\sig\to\sig/\sig_\infty$ is dimensionless and $>1$. This solution assumes 
that the singularity of the field is always at the tip, which is consistent 
with a quasi-static picture. Consider, however, a situation wherein the dynamic
response constrains the tip to overtake the density waves that adjust the
field. In this situation the singularity in the stress field {\it does not}
coincide with the location of the tip and the tip's stress drops to below the
static value. The difference between the static and dynamic stresses at the tip
depends on the tip's velocity, $v=dl/dt$ and the propagation history. The
dynamic stress is found from $\Iii$ by putting $\zeta=(l-ct)\Theta(l-ct)$,
where $l$ is the tip's position and the step-function, $\Theta$, ensures
that when the shear wave catches up the tip stress stays at the
static value. When $\Theta=0$ the tip stress diverges as expected and 
traditional quasi-static solutions apply \ref\kos{B.V.Kostrov, Appl. Math. 
Mech. (PMM) {\bf 30}, 1241 (1966)}. Focusing on non-quasi-static propagation, 
I assume $\Theta=1$ during the entire growth. When $\Theta$ alternates 
between 0 and 1 one simply pieces the solutions together.
Taking the time derivative of $\Iii$ we have
\eqn\Iiii{\dot\sig = -\dot\zeta a^2 / [\zeta(\zeta+2a)]^{3/2} \ .}
Using $\Iii$, we can invert relation $\Iiii$:
\eqn\Iiiia{\dot\zeta / a = -\dot\sig / \left(\sig^2 - 1\right)^{3/2} \ .}
We define $\dot\zeta=(v-c)$ as $\dot\zeta\equiv cu$, where u is a reduced
velocity. Upon substitution in $\Iiiia$ we can readily solve the equation:
\eqn\Iiv{t-t_0 = -{a\over c}\int^{\sig(t)}_{\sig(t_0)}
{{ds}\over{(s^2-1)^{3/2}u(s)}} \ .}
The kinetics are thus determined by the response function through the
stress-dependence of $u(\sig)$. Relation $\Iiv$ is the bare result of this
report. It is an exact derivation from the Yoffe solution. It gives the 
general time dependence of the stress at the crack tip.
Once the stress history is found from this relation one substitutes it in
$v(\sig)$ to obtain the velocity history. We now proceed to analyse the
consequences of this result, assuming  the qualitative response shown in Fig.
1. It is convenient to classify the behavior in terms of the ratio
$\lambda=c/v_l$. The reason is that, as is shown below, the mode of propagation
depends only on this ratio.

{\bf \ } \pmb{$\lambda>1$}: The point ($\sig(c),\ c$) is on the upper branch of
$v(\sig)$. Suppose that initially $\sig<\sig_l$. The velocity is momentarily
zero (or very low) and the tip stress builds up to $\sig_h$. At this stage the
system `jumps' to the upper branch and fast motion ensues. From relation $\Iiv$
we see that for $\sig_h>\sig(c)$ ($<\sig(c)$) the stress will decrease
(increase) until $\sig$ converges to $\sig(c)$ whereafter the tip propagates at
a velocity $c$ and a fixed distance ahead of the density waves. Thus
($\sig(c),\ c$) is a {\it fixed point} of the equation of motion. The behavior
at the vicinity of this point can be found by linearization of relation $\Iiv$:
\eqn\Iv{\left|\sig - \sig(c)\right| \approx C e^{-\gamma t} \halfspace ;
\halfspace \gamma\equiv{{\left[\sig(c)^2-1\right]^{3/2}}\over
a}\left({{dv}\over{d\sig}}\right)_{\sig(c)} \ .}
Since $v(\sig)$ near $\sig(c)$ is smooth and positive $\gamma$ is regular and
positive and the fixed point is {\it stable}, namely, a steady state
propagation at a limiting velocity $c$ is a stable fixed point of the dynamics.
A typical such history of $v$ is shown in fig. 2. An interesting implication of
this result is that the experimentally observed limiting crack velocities give
in fact the value of $c$ and hence the local stress relaxation rate. This
suggests a check of this model by comparing the limiting velocity to the speed
of sound in the PZ. It is intriguing to note that even in the absence of a
global energy balance criterion the crack velocity converges to the SWS, albeit
the local value, $c$.
Observations that $v$ increases very slowly with $K$ \ref\flucv{J.W.Dally and
A.Shukla, Mech. Res. Comm. {\bf 6}, 239, (1979); T.Kobayashi and J.T.Metcalf in
"Crack Arrest Methodology and Applications" ASTM-STP {\bf 711}, 128 (1980);
J.F.Kalthoff, in Proc. of "Workshop on Dynamic Fracture" CALTECH, 11 (1983);
K.Ravi-Chandar and W.G.Knauss, Int. J. Fract. {\bf 26}, 141 (1984)}$\ $ in this
regime indicate a small value of $dv/d\sig$ along the upper branch. In view of
the present analysis, this agrees with the reported velocity behavior
immediately after crack initiation \rki\hyst\df. Another check of this picture
can be suggested: In some experiments a drop in the stress has been measured
after crack initiation \ref\deccel{J.F. Kalthoff {\it et al.}, in {\it Fast
Fracture and Crack Arrest} ASTM STP 627, 161 (1977)}. This suggests that in
those systems $\sig(c)<\sig_h$, a conclusion that can be checked by independent
methods as a test of this analysis.

{\bf \ } \pmb{$\lambda<1$}: To analyse this case, let us assume again that
initially the tip stress is lower than $\sig_l$. From relation $\Iiv$ the
stress will increase until it reaches $\sig_h$, whereupon the crack will start
propagating as for $\lambda>1$. The velocity and the stress will then gradually
decrease. Since $c<v_l$ is not a point on the upper branch the system cannot
settle into a steady state as before and at $\sig_l$ it flips back to the lower
branch. There the crack halts momentarily, the stress at the tip builds up
again to $\sigma_h$ and the cycle repeats itself. This is a {\it relaxation
cycle} whose period is found from $\Iiv$:
\eqn\Cii{\tau = {a\over c} \int_{\sig_l}^{\sig_h} {{1/u_{ub}(s) -
1/u_{lb}(s)}\over{(s^2-1)^{3/2}}} ds \ ,}
where $u_{ub}>0$ and $u_{lb}<0$ are, respectively, the values of $u$ along the
upper and lower branches. When the velocity vanishes along the lower branch
$u_{lb}=-1$. A typical velocity history in this case is also shown in fig. 2.

{\bf \ } \pmb{$\lambda=1$}: This marginal case is sensitive to the value of
$\partial v/\partial\sig$ at $\sig_l$. If the derivative is regular one can
easily see that the analysis is the same as for $\lambda>1$. The only
difference is that $\sig$ can only approach $\sig(c)=\sig_l$ from above because
for $\sig<\sig_l$ the only motion is up the lower branch. If $\partial
v/\partial\sig$ diverges at $\sig_l$ the behavior depends on the detailed form
of the divergence. For illustration, consider the form
\eqn\Ivi{v = v_l
\exp{\left[\alpha\left({\sig\over{\sig_l}}-1\right)^\nu\right]} \to u \sim
Const. + \left(\sigma - \sigma_l\right)^\nu\ ,}
with $0<\nu<1$. The behavior near the fixed point is found by using $\Ivi$ in
Eq. $\Iiv$,
\eqn\Ivii{\sig - \sig(c) \sim \left(\tau_0 - t\right)^{1/(1-\nu)} \halfspace
; \halfspace u \sim \left(\tau_0 - t\right)^{\nu/(1-\nu)}\ ,}
where $\tau_0>t$ is a constant. Now the propagation rate converges to $c$ as a
{\it power law} rather than exponentially. It should be noted that this
propagation mode is sensitive to small fluctuations that can easily flip the
system to the lower branch. A small noise in this case will give rise to a
quasi-periodic motion similar to that discussed below.

{\bf Intermittency and quasi-periodicity}: Since the PZ is inhomogeneous one
expects fluctuations in the local properties which may well go beyond the
effective medium assumption. For simplicity, let us restrict the discussion
only to fluctuations in the tip stress during propagation,
$\sig=\sig_0(t)-\eta(t)$, where $\sig_0$ is the stress in the absence of noise.
This corresponds to a situation where the crack encounters microvoids of 
varying
sizes along its path. Upon association of a microvoid to the crack the tip
stress drops momentarily with the drop depending on the microvoid's size.
The fluctuations in microvoid sizes give rise then to noise in the tip stress.
A fluctuation during steady-state propagation that reduces $\sig$ by more than
$\delta=\sig(c)-\sig_l$ (see fig. 1) flips the system to the lower branch. The
system then has to go through the process of stress increase to $\sig_h$, jump
to the upper branch and convergence to $\sig(c)$ again. The time that this
process takes depends on the original fluctuation, $\eta>\delta$, and can be
found by applying $\Iiv$ to the motion along the two branches:
\eqn\Ciii{T(\eta) = {a\over c}\left[ \int_{\sig(c)}^{\sig_h}
{u_{ub}(s)^{-1}{ds}\over{(s^2-1)^{3/2}}} - \int_{\sig(c)-\eta}^{\sig_h}
{u_{lb}(s)^{-1}{ds}\over{(s^2-1)^{3/2}}}\right] \ .}
It is the occurrence frequency of the flips between the branches which
determines to a large extent the observable behavior. This frequency depends
both on the noise characteristics and the value of $\delta$. The stochastic
velocity behavior can be obtained from the statistics of $\eta$ by using
relation $\Ciii$. For example, the probability density of $T$, $P(T)$, can be
found from that of $\eta$, $P_0(\eta)$, by inverting relation $\Ciii$ to obtain
$\eta(T)$ and substituting in
\eqn\Cv{P(T) = P_0\left(\eta(T)\right)\ (d\eta/dT)\ .}
{}From $\Ciii$ one can also find the effects of various forms of the noise
temporal correlations, $<\!\eta(t)\!\eta(t^\prime)\!>$ on the velocity history.
A detailed analysis of the statistics, including the explicit dependence on the
distribution of microvoid sizes, will be reported shortly. Here I only point
out a few intriguing consequences for $\lambda\!>\!1$ and $\sig(c)\!<\!\sig_h$:
First, a low occurrence frequency (i.e., $<\!1/\tau$) of $\eta\!>\!\delta$ 
leads
to an {\it intermittent} behavior wherein the tip is `knocked' occasionally
from the steady state and then returns to it only to be knocked out of it again
at a later time. A plot of such a history is shown in fig. 3. Second, a very
high occurence frequency of fluctuations $\eta\!>\!\sig_h\!-\!\sig_l$ gives
rise to a quasi-periodic behavior as follows: As the tip stress builds up along
the lower branch to $\sig_h$ the system flips to the upper branch. A
fluctuation then immediately knocks the system back to the lower branch, not
allowing it to settle into the steady state. The mean period will be then close
to the time spent on the lower branch, namely,
$\approx\!(a/c)\!\left[(1\!-\!1/\sig_l^2)^{-1/2}\!-
\!(1\!-\!1/\sig_h^2)^{-1/2}\right]$. Thus the microvoid size distribution
determines the observable behavior by governing the statistics of $\eta$. Since
this distribution also plays a major role in determining the roughness of the
fracture surfaces my analysis can rigorously link roughness
measurements to the velocity history.

To summarize, a minimal theoretical model has been proposed to explain the rich
behavior observed in crack propagation in amorphous and polymeric materials.
The theory is a direct consequence of the observations of slow relaxation rates
of the tip stress, the occurence of different initiation and arrest stresses,
and the deduced qualitative form of $v(\sig)$. The mode of propagation has been
found to depend only on one material parameter, $\lambda=c/v_l$. For
$\lambda>1$ the propagation speed saturates to a limiting value, while for
$\lambda<1$ it oscillates periodically. Noise gives rise to a spectrum of
behavior ranging from intermittent to quasi-periodic propagation. The model
explains naturally recent observations of oscillations in polymeric
materials and measurements for its validation have been suggested.
It predicts that the low-noise steady-state growth rate is exactly $c$, the 
wave speed in the PZ, and should be possible to test experimentally. Low 
$\lambda$ is expected to correspond to high disorder and vice versa. So by 
manipulating the disorder one may tune $\lambda$.
I should remark that the effective continuum approximation of the PZ probably
breaks down for too broad a distribution of microvoid sizes, and a 
statistical treatment is more adequate. E.g., for
propagation velocities of order $c \sim 500$m/s and given the fact that
currently observations are limited to times of $\mu$sec and higher, the
effective continuum assumption should hold when microvoids are smaller than
$500\mu$m. Microvoids do
not usually reach such sizes in polymeric materials and therefore this model
should do a good job explaining experimental observations \osc $\ $in these
systems. A complementary statistical analysis for broad microvoid
distributions is currently under way and will be reported shortly. Finally,
many ramifications of this model remain to be explored: the effects of noise
correlations $<\eta(t)\eta(t^\prime)>$ on the dynamics, the effects of
realistic distribution of void sizes, and the implications of the statistics 
and velocity history on the surface roughness, to name a few.

\bigskip
\centerline {\bf FIGURE CAPTIONS}

\item {1.} A generic plot of $v(\sig)$.

\item {2.} Typical velocity histories in the steady-state (solid) and 
periodic (dashed) regimes.

\item {3.} A typical velocity history in the intermittent regime.

\bigskip

\vfill\immediate\closeout\rfile
\baselineskip=18pt\centerline{{\bf REFERENCES}}\bigskip\frenchspacing%
\input refs.tmp\vfill\eject\nonfrenchspacing

\includegraphics{v-sigma11.ps}

\includegraphics{v-t2.ps}

\includegraphics{vinter-t2.ps}

\bye